\documentstyle[epsfig]{aipproc}
\title{A Gamma-Ray Burst Bibliography, 1973-1999}
\author{Kevin Hurley }
\address{UC Berkeley Space Sciences Laboratory \\
Berkeley, CA 94720-7450
}
\begin{document}

\maketitle

\begin{abstract}
On the average, one new publication on cosmic gamma-ray bursts
enters the literature every day.  The total number now exceeds 4100.
I present here a complete bibliography which can be made available
electronically to interested parties.
\end{abstract}

\section{Introduction}
I have been tracking the gamma-ray burst literature for about the 
past twenty-one years, keeping the 
authors, titles, references, and key 
subject words in a machine-readable form.  The present 
version updates previous ones reported
in 1994, 1996, and 1998 \cite{hurley94,hurley96,hurley98}.  In 
its current form, this information is 
in a Microsoft Word 97 
``doc'' format.  My purpose in 
doing this was first, to be able to 
retrieve rapidly any articles on a 
given topic, and second, to be able 
to cut and paste references into 
manuscripts in preparation.  The 
following journals have been 
scanned on a more or less regular 
basis starting with the 1973 
issues:\\
\\
Advances in Physics* \\
Annals of Physics* \\
Astronomical Journal*\\ 
Astronomische Nachrichten*\\ 
Astronomy and Astrophysics (letters, main journal, and supplement series)*\\ 
Astronomy and Astrophysics Review* \\
Astronomy Letters* (formerly 
Soviet Astronomy Letters)\\ 
Astronomy Reports*(formerly 
Soviet Astronomy) \\
Astrophysical Journal (letters, 
main journal, and supplements)*\\ 
Astrophysical Letters and 
Communications \\
Astrophysics and Space 
Science* \\
ESA Bulletin*\\ 
ESA Journal* \\
IEEE Transactions on Nuclear 
Science* \\
Journal of Astrophysics and 
Astronomy* \\
Monthly Notices of the Royal 
Astronomical Society*\\ 
Nature \\
Nuclear Instruments and 
Methods in Physics Research
Section A* \\
Observatory* \\
Physical Review (main journal A 
and letters)* \\
Proceedings of the Astronomical 
Society of Australia*\\ 
Publications of the Astronomical 
Society of Japan* \\
Publications of the Astronomical 
Society of the Pacific*\\ 
Reports on Progress in Physics*\\ 
Science* \\
Scientific American\\
Sky \& Telescope\\
\\
The asterisks indicate journals 
which are scanned using the on-line version of Current Contents.  
In addition, the following journals 
have been scanned, but in many 
cases less regularly, particularly in 
the past:\\
\\ 
Annals of Geophysics \\
Astrofizika \\
Bulletin of the American 
Astronomical Society\\ 
Bulletin of the American 
Physical Society \\
Chinese Astronomy \\
Cosmic Research \\
Journal of Atmospheric and 
Terrestrial Physics\\ 
Journal of the British 
Interplanetary Society\\ 
Journal of the Royal 
Astronomical Society of Canada\\ 
Progress in Theoretical Physics \\
Solar Physics \\
Soviet Physics \\
\\
The above lists are not 
exhaustive.  For example, where 
theses, newspaper articles, or internal reports have 
come to my attention, I have 
included them, too.  To be included, 
an article had to have something to 
do with gamma-ray burst theory, 
observation, or instrumentation, or 
be closely related to one of these 
topics (e.g., merging neutron 
stars, AXPs, SGRs, the Bursting Pulsar), and must have been 
published in some form.  With only a few 
exceptions, preprints which were 
never published have not been 
included. 
 
\section{Organization of the Bibliography}

The overall organization is 
chronological by year.  Within a 
given year, articles published in 
journals are listed first, in 
alphabetical order by first author.  
Then come theses and conference 
proceedings articles.  The latter 
are listed in the order in which they 
appear in the proceedings.  

\begin{figure}[tbh]
\centerline{\epsfig{file=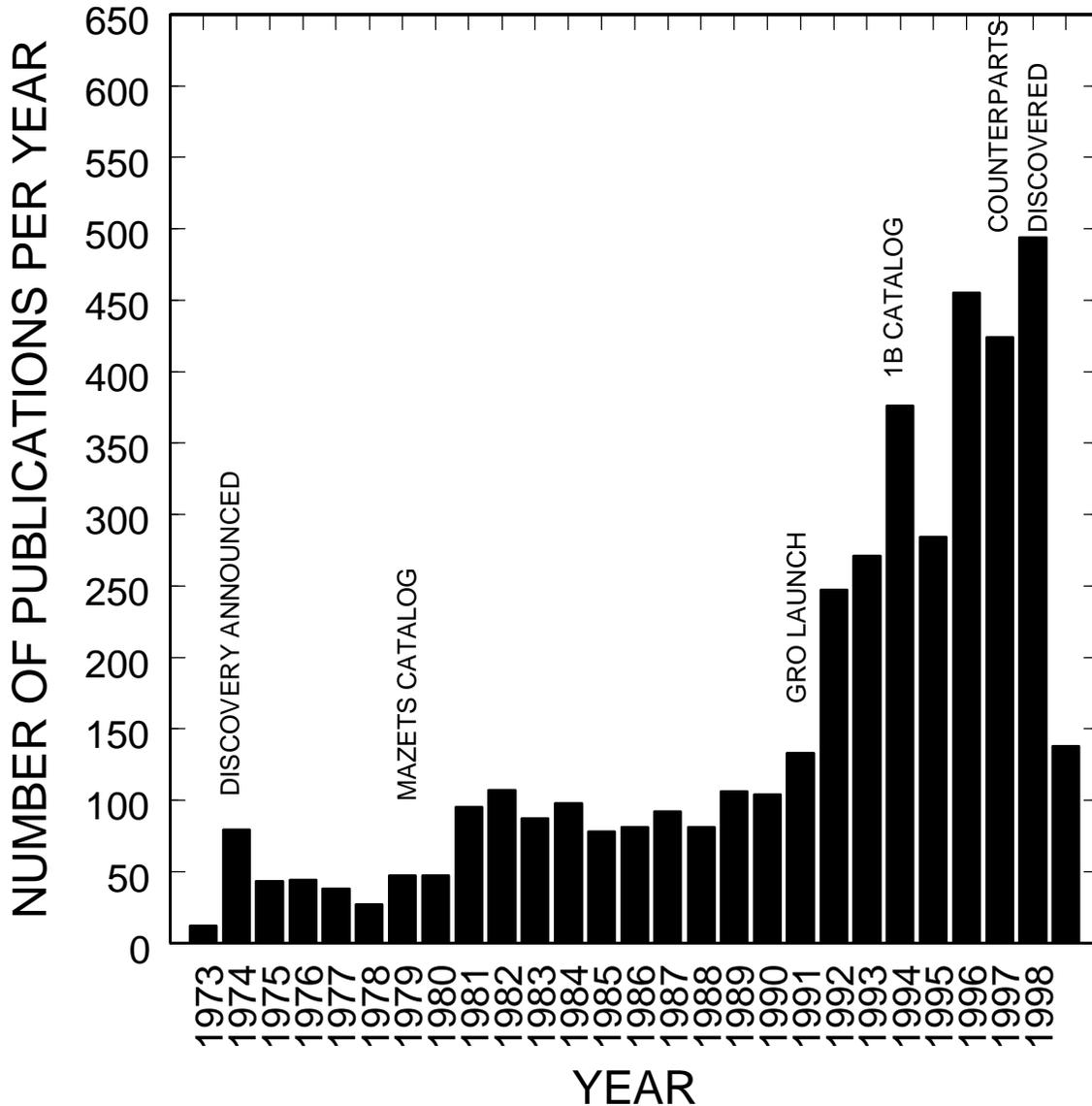,width=15 cm}}
\caption{Number of gamma-ray burst publications as a function of year.  The cutoff 
date is mid-1999.}
\label{fig1}
\end{figure}

\begin{figure}[h!]
\centerline{\epsfig{file=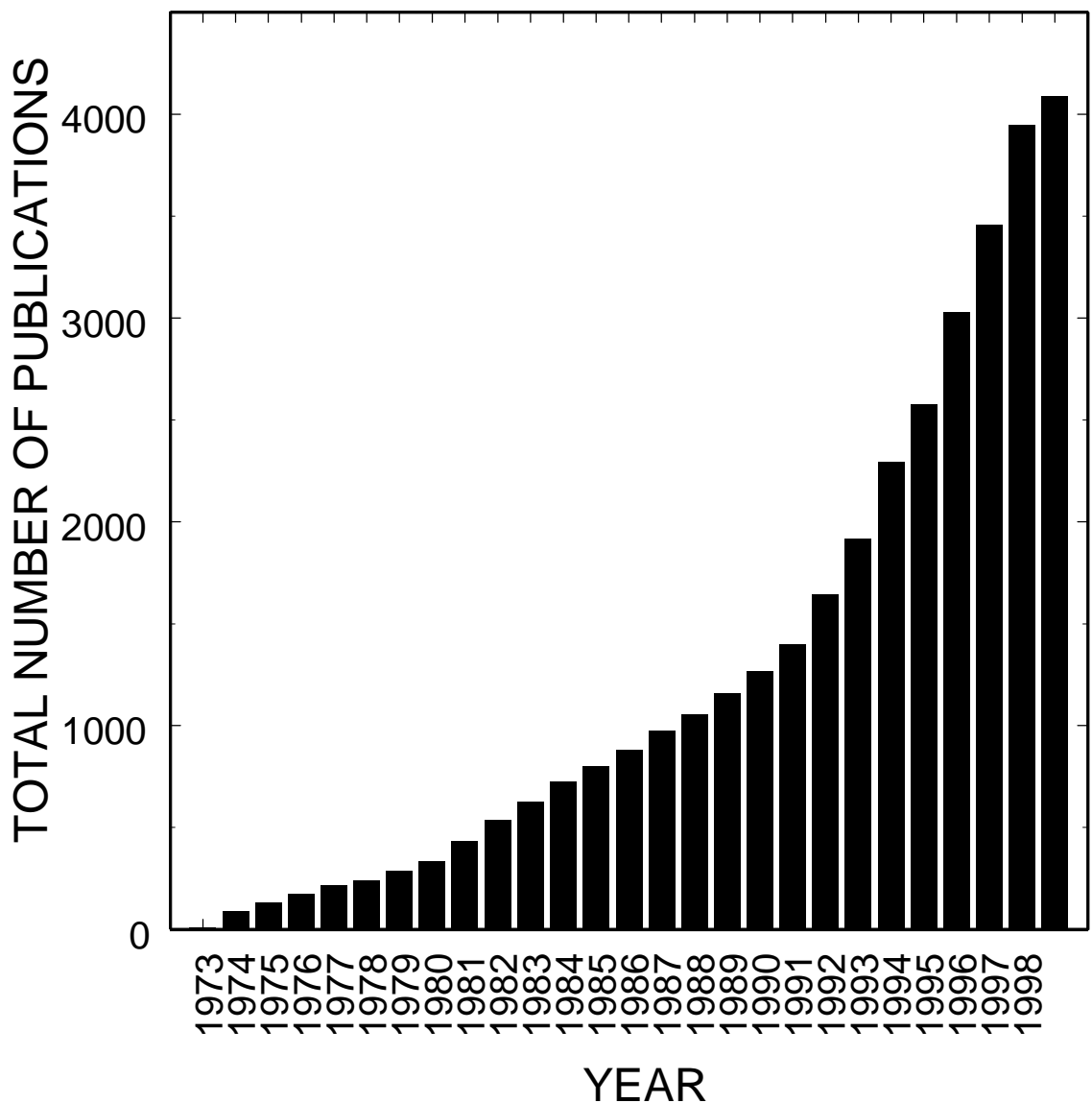,width=15 cm}}
\caption{The cumulative number of publications by year. }
\label{fig2}
\end{figure}

The entries are numbered 
consecutively, so that paper copies 
which are kept on file can be 
retrieved quickly.  However, to 
avoid having to renumber this 
entire file when a new article is 
added, numbers are skipped at the 
end of each year and reserved for 
later inclusion.  The complete 
author list follows, as it appears in 
the journal, along with the title, 
journal, volume number, page 
number, and year.  A line 
containing key words follows this.  
These are generally not the same 
key words as the ones listed in the 
journal, nor are they taken from the 
title or any particular list.  Rather, 
they are meant to reflect the true 
content of the article, and provide 
a list of machine-searchable 
topics.  In general, however, key 
words have not been included for 
conference proceedings articles.

\section{A Few Interesting Statistics}

The number of articles 
published each year since 1973 is 
shown in figure
 \ref{fig1}.  Starting with a 
modest article per month  in 1973, 
it began to exceed one per day in 
1994.  Several milestones are indicated as the
probable causes of sudden increases in the publication rate.
The apparent decreases in the rates in 1995 and 1997 are in fact
due to a 2 year periodicity in the publications caused by the influx
of a large number of articles from the Huntsville Workshop proceedings.
In keeping with this publication rate, the bibliography
is updated on an approximately daily basis. 
Note that there are still about as 
many papers published as there 
are gamma-ray bursts.  The 
cumulative total is shown in figure \ref{fig2}.

The sheer volume of the 
literature has necessitated the 
development of a program which 
can search for and extract 
particular articles.  I have written 
such a program in Microsoft Word 
Basic (a variant of the BASIC 
programming language).  It allows 
one to extract all articles between 
two dates whose entries contain a 
particular key phrase and write it 
to a separate file.

\section{Availability}

The IPN web site ssl.berkeley.edu/ipn3/index.html contains a version of this bibliography.  
More up-to-date versions  
in plain ASCII, ``doc'', and ``rich text'' (rtf) formats can be made available 
to interested parties as time 
permits.    Please contact me at
\begin{center} 
khurley@sunspot.ssl.berkeley.edu\\
\end{center} 
to request copies, and indicate 
your preference for the format.  I would appreciate it 
if users would communicate errors 
and omissions to me.

This work was carried out under 
JPL Contract 958056 and CGRO 
guest investigator grant NAG5-7810.

\end{document}